\documentclass[twocolumn,pre,floatfix,showpacs]{revtex4}
\usepackage{dcolumn}
\usepackage{graphicx}
\usepackage{graphicx,amssymb,amsmath}
\usepackage{natbib}

\begin{document}

%
%
%
%
%
%
%
%

\title{Diffusion limited reactions in confined environments}
\author{Jeremy D.\ Schmit}
\email[]{schmit@brandeis.edu}

\author{Ercan Kamber}
\email[]{ekamber@brandeis.edu}


\author{Jan\'{e} Kondev}
\email[]{kondev@brandeis.edu}
 \affiliation{Department of Physics,
Brandeis University, Waltham MA 02454 USA.}

\begin{abstract}
We study the effect of confinement
on diffusion limited bi-molecular reactions within a lattice model
where a small
number of reactants diffuse amongst a much larger number of inert
particles.  When the number of inert particles is held constant
the rate of the reaction is slow for small reaction volumes
due to limited mobility from crowding, and for large reaction volumes
due to the reduced
concentration of the reactants.  The reaction rate proceeds fastest
at an intermediate confinement corresponding to volume fraction
near 1/2 and 1/3 in two and three dimensions, respectively.  We
generalize the model to off-lattice systems with hydrodynamic coupling
and predict that the optimal reaction rate for mono-disperse colloidal
systems occurs when the volume fraction is $\sim 0.18$.  Finally, we
discuss the application of our model to bi-molecular reactions inside
cells as well as the dynamics of confined polymers.
\end{abstract}

\maketitle
It is a somewhat surprising fact that the total concentration of protein
within a cell rivals that within a protein crystal\cite{Fulton:82}.
This highly
crowded environment plays an important role in dynamical processes, such
as rates of chemical reactions, and thermodynamic properties,
such as chemical equilibria, observed
{\it in vivo}\cite{Zimmerman:93}.  While the volume fraction of
macromolecules within the cell may exceed 30\%,
there may be just a few copies of a given protein corresponding to a
concentration of a few nanomolars for a cell of volume
$1 {\rm \mu m}^3$\cite{Zimmerman:91,Hall:03}.  This
is in stark contrast to {\it in vitro} experiments where the reactants
are present at a relatively high concentration with a negligible level of
crowding molecules.  Therefore, caution is required when interpreting {\it in vivo}
biochemical experiments as it is not always obvious what effect the presence of
crowding molecules will have.  For example, the rate of a reaction may be
increased if the crowding favors a compact transition state, decreased if
the reaction is diffusion limited, or unaffected if the reactants are small
compared to the crowding species\cite{Minton:06}.

In this letter we study the effect of crowding on reaction rates in a
finite system with a fixed number of particles.  A biological cell
represents such a system.  We find that the presence of non-reactive
particles leads to a non-monotonic reaction rate as the volume of the
system is changed.  This non-monotonicity is the result of
two distinct dynamical regimes.  When the system volume is very large the effect of
the crowding particles is negligible.  Therefore, the rate of reaction is
dictated by the time required for the reactants to diffuse the mean
separation between reactive particles which increases with the
system size.  We will call such systems
``concentration limited''.  At the opposite extreme is the case where
the system volume is very close to the sum of the total volume of the
reactant and crowding particles.  In this case the separation between
reactants may be
quite small, but the reaction proceeds slowly because the high density
of crowding particles severely impedes the diffusion of the reactants.
Such systems are ``crowding limited''.  Between these two limits there
is an optimal volume at which the reaction proceeds the fastest.

The competition between the concentration limited and crowding
limited regimes may be understood through the following simple
argument.  The flux of particles at the surface of an absorbing
sphere is $4 \pi a c_r D$, where $a$ is the radius of the sphere,
and the concentration of particles far from the sphere,  $c_r$, is
inversely proportional to the volume of the system $c_r\propto
R^{-3}$\cite{Berg:83}. The effect of crowding particles that do not
react with the absorbing sphere may be included through a rescaling
of the diffusion constant $D$ provided the distance the reactant
particles must travel is large compared to the mean spacing between
crowding particles.  For the case of a lattice-based system, a
mean-field diffusion constant (which is exact when the crowding
particles move much faster than the reactants) is achieved by
multiplying the ``bare'' diffusion constant $D_0$ by the success
rate for a particle move.  This results in a diffusion constant of
the form $D\sim D_0 p$ where $p=(1-c)$ is the probability that the
neighboring site for an attempted move is unoccupied, and $c$ is the
number density of particles on the lattice.  Therefore, the flux at
the absorbing surface scales as $a (1-N b^3 R^{-3})/R^3$, where $N$
is the total number of particles in the system and $b$ is the
lattice constant. This expression has a maximum when the system size
is such that the particle density is $1/3$.

In order to go beyond this simple argument, we have explored
the transition from concentration limited reactions
to crowding limited reactions using an analytically tractable model with
two reactant particles that react instantaneously upon
contact.  With this simplification, the reaction ``rate'' is just the
inverse mean first passage time for the particles to find each other.
If we make the further simplification of holding one of the reactants
fixed, the mean first passage time is given by
\begin{equation}
\bar{\tau}=\frac{1}{V}\int_V \tau(\vec{x}) d^d \vec{x}
\label{eq:tauavg}
\end{equation}
where the integral is over the $d$-dimensional volume of the system.
$\tau(\vec{x})$, the average
time for the mobile reactant at $\vec{x}$ to reach the stationary target,
satisfies the equation \cite{Redner:01}
\begin{equation}
D \nabla^2 \tau(\vec{x}) = -1.
\label{eq:tauX}
\end{equation}
Eq. \ref{eq:tauX} is subject
to a reflecting boundary condition at the system periphery and an absorbing
boundary condition at the surface of the stationary reactant.

\begin{figure}[ht]
\vspace{0.6 cm}
\begin{center}
\includegraphics[angle=0,scale=0.2]{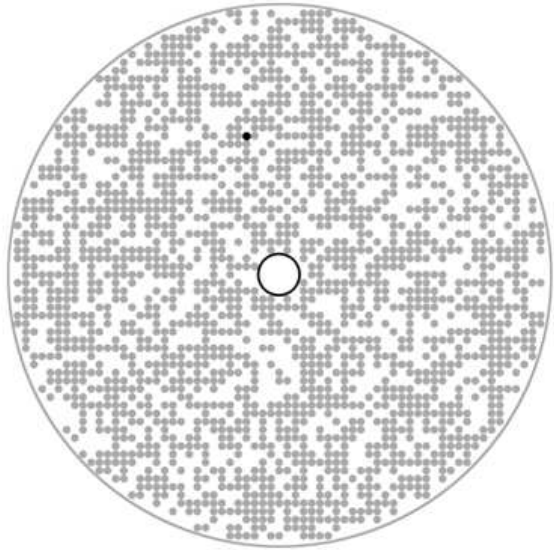}
\end{center}
\caption{
\label{fig:snap}
Snapshot from the simulation described in the text.
The stationary target is indicated by the black circle
in the center, and the mobile reactant
is the black dot halfway between the target and the top of the
outer circle.  Also shown are the 2000 inert crowding
particles (grey).
}
\end{figure}

\begin{figure}[ht]
\vspace{0.6 cm}
\begin{center}
\includegraphics[angle=0,scale=0.6]{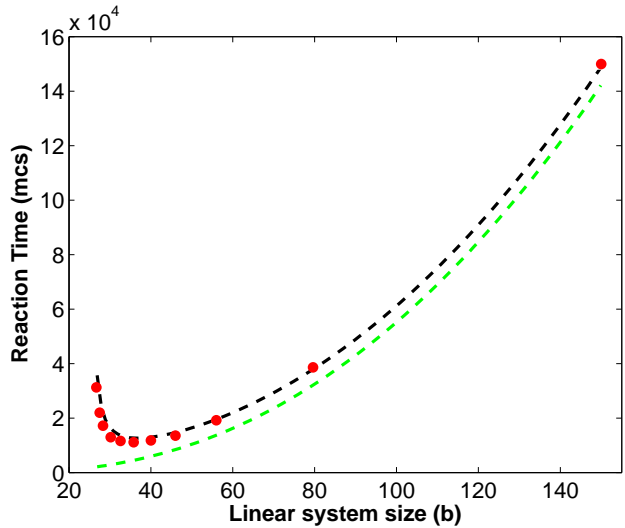}
\end{center}
\caption{ \label{fig:freeparticles} (color online) Comparison of the
reaction time predicted by Eq. \ref{eq:tauavg} (black dashes) to the
simulation data (red dots).  The green dashes indicate the reaction
time predicted in the absence of crowding particles.  Error bars
(not shown) are smaller than the size of the dots.}
\end{figure}

If we place the fixed reactant at the center of a spherical box,
as shown in Fig. \ref{fig:snap}, Eq. \ref{eq:tauavg} is exactly solvable
with the result (in two and three dimensions)
\begin{eqnarray}
\label{eq:twoDtau}
\bar{\tau}_{2D}&=&\frac{1}{2 D(c)(R^2-a^2)}\left(R^4 \ln\left(\frac{R}{a}\right) -\frac{3 R^4}{4} + R^2 a^2 -\frac{a^4}{4}\right) \\
\label{eq:threeDtau}
\bar{\tau}_{3D}&=&\frac{6}{D(c)(R^3-a^3)}\left(\frac{R^6}{3 a} -\frac{3 R^5}{5} + \frac{R^3 a^2}{3} -\frac{a^5}{15}\right).
\end{eqnarray}
Here $a$ is the radius of the stationary target, and $D(c)$ is the
concentration dependent diffusion constant.  In the spirit of the
mean-field argument introduced earlier, we employ a lattice model
and utilize an excellent approximation for the self-diffusion of a
lattice gas derived by van Beijeren and Kutner
\cite{VanBeijeren:85}.  For reactant and crowding particles with
equal mobilities on a square lattice, the self-diffusion constant
takes the form
\begin{equation}
D(c)=\frac{\Gamma b^2}{8}\left(\sqrt{4(1-c)+c^2(\pi-1)^2}-c(\pi-1)\right),
\label{eq:diffVB}
\end{equation}
where $b$ is the lattice spacing, $\Gamma$ is the attempt rate for
particle moves, and $c$ is the ratio between the total number of
mobile particles
to the number of accessible sites $c\simeq b^2(N+1)/(\pi(R^2-a^2))$.
>From Eqs. \ref{eq:twoDtau} and \ref{eq:diffVB} we find an expression
for the reaction time which is plotted in
Fig. \ref{fig:freeparticles}.

We have preformed Monte Carlo simulations of this model consisting
of one reactant and $N$ inert particles confined to a two
dimensional circular box of radius $R$ with a circular target of
radius $a$ placed at the center, as shown in Fig. \ref{fig:snap}.
Each run of the simulation begins with a random configuration of the
reactant and crowding particles and ends when the reactant reaches
the center target at which point the first passage time is recorded.
The average reaction time is determined from 10,000 such runs.

For the purposes of our simulation we define the box and the target
to include all sites on the lattice with a distance to the origin
less than or equal to $R$ and $a$, respectively.  Although the
definition of a circular box on a square lattice is somewhat
cumbersome, this geometry is advantageous relative to square or
rectangular boxes because the circular box allows for non-integer
adjustments of the linear box dimension with minimal change to the
overall geometry.  This fine-tuning of the box size greatly
increases the number of data points that can be collected in the
crowding limited regime.

The simulation used $N=2000$ crowding particles which, like the
single mobile reactant, each occupied a single lattice site.  The
central target had a larger size $a=3$ to minimize the discreetness
effect of the lattice when comparing to the continuum theory given
by Eq. \ref{eq:tauavg}.  The radius of the confining box was varied
from $R=150$ to $R=26.7$ lattice spacings, corresponding to
densities from $c=0.03$ to $c=0.90$.  The results are shown in Fig.
\ref{fig:freeparticles}.

The simulation results show a minimum in the reaction time very near
the minimum of 36.9 predicted by Eq. \ref{eq:twoDtau} (corresponding
to 47\% of the sites being occupied).  The reaction times diverge
sharply when the box radius becomes less than 29 lattice sites, or
the concentration exceeds $\sim 75 \%$.  At large box radii the
reaction times increase as $R^2$ and approach the reaction rate in
the absence of crowding particles (green line).  As shown in Fig.
\ref{fig:freeparticles} the difference between the crowded and
uncrowded reaction times is nonzero at all system sizes. This can be
explained by noting that while the probability that a given site is
occupied scales as $R^{-d}$, the required number of steps scales
like $R^d$.  The result is that the number of time steps where the
particle is unable to move due to crowding remains nearly constant
as the system size is increased.

\begin{figure}[ht]
\vspace{0.6 cm}
\begin{center}
\includegraphics[angle=0,scale=0.45]{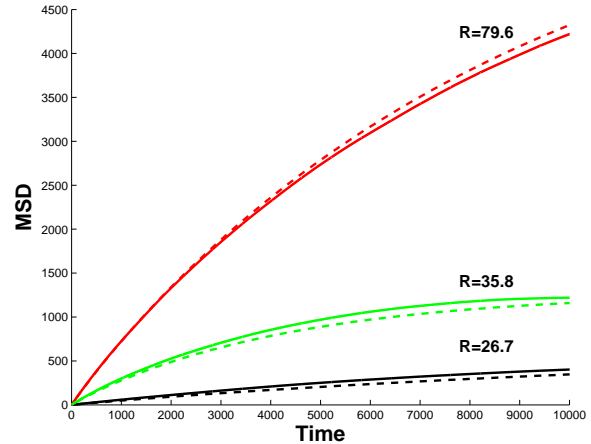}
\end{center}
\caption{ \label{fig:MSD} (color online) The mean mean squared
displacement of the mobile reactant measured in our simulation
(solid lines) compared to the calculated mean squared displacement
for a random walker in a circle(dashed lines).}
\end{figure}

The reaction times in Fig. \ref{fig:freeparticles} deviate from the
theoretical curve at system sizes below $50b$ with a maximum error
of 39\% at the highest concentration simulated.  This discrepancy is
due to a non-diffusive, logarithmic correction to the mean squared
displacement (MSD) in the lattice gas system\cite{Saxton:87}. This
correction allows the reactant to sample its immediate surroundings
more efficiently than a purely diffusive particle.  The effect of
the correction term can be seen in Fig. \ref{fig:MSD} where we
compare the MSD of the reactant in our simulations to the expected
MSD (as calculated from the Green's function) for a random walker in
a circle with diffusion constant given by Eq. \ref{eq:diffVB}.  At
the largest system size, when the particle density is 0.1, the
agreement between the simulation and calculated MSD is better than
3\% at all times. At $R=35.8$ and $R=26.7$, where the particle
density is 0.5 and 0.1 respectively, the discrepancy between the
observed and calculated MSD is 8\% and 18\% respectively at short
times and shrinks at later times as the MSD saturates.

Our results can be extended to the more physically relevant three
dimensional, off-lattice system with a few modifications. We recall
that hard sphere systems undergo a glass transition at a volume
fraction $\phi_c \sim 0.58$, which is less than the close packing
density\cite{Weeks:00}.  Although the system is not completely
frozen, the relevant time scales diverge sharply.  Therefore, we
restrict our analysis to densities below $\phi_c$.  In the fluid
regime, for $\phi<\phi_c$, the self diffusion constant is modified
not only by the short range excluded volume interactions, but also
by long range hydrodynamic coupling mediated by the
solvent\cite{Tokuyama:95}.  This latter effect is not included in
our lattice model due to the lack of solvent.  Using the
self-diffusion constant derived in reference \cite{Tokuyama:95}
(equation 4.34), together with Eq. \ref{eq:threeDtau}, we predict
that the optimal reaction time for mono-disperse, hard-core,
Brownian spheres occurs when the system size is such that the volume
fraction is $\sim 0.18$.

Experimentally, a crowded reaction with adjustable volume could be
realized using microfluidic
techniques.  In this case the tracer and crowding particles would be
confined to a micro droplet reaction vessel whose volume could be controlled
through osmotic gradients across a semi-permeable barrier\cite{Shim:07}.
This system mimics {\it in vivo} experiments that show increased tracer
particle diffusion in osmotically swollen cells\cite{Kao:93,Swaminathan:97}.

In most systems of interest the target reactant would be mobile rather
than stationary.  One expects that the ratio of the
reaction time for the stationary target to the reaction time for two
mobile reactants in a fixed box size should be independent of the number
of crowding particles
present, as the crowding particles merely rescale the diffusion time.
Our simulations support this intuition, and we find that the ratio is
$\sim 1.6$ and $\sim 2.0$ in two and three dimensions respectively, with
corrections on the order of $a/R$.

In the context of these theoretical results it is natural
to ask to what extent do cells optimize biochemical
reaction rates by adjusting their size?
At first glance it would seem unlikely that cell volume
would be a useful parameter for the cell to use to regulate reaction rates
due to the large number of reactions that occur simultaneously.
Furthermore, differential protein expression rates during the cell
cycle could modify the overall protein concentration and thereby alter
the cytoplasmic diffusion rate\cite{Elowitz:99}.  However,
it is well known that the cell is more complicated than a ``bag
of enzymes,'' and therefore, the potential exists for the cell
to compartmentalize reactions in such a way that the
volumes of the compartments are individually tunable.
For example, the digestion
of a pathogen within a macrophage occurs within a vesicle created
by endocytosis.  The volume of this vesicle is, in principal,
adjustable by the amount of membrane used during vesicle creation.
Similarly, other membrane-bound organelles such as golgi, endoplasmic
reticulum, mitochondria, and the nucleus could be individually
adjusted to optimize reactions occurring within.  This is consistent
with the finding that diffusion rates in the mitochondria and
endoplasmic reticulum can differ substantially from the
cytoplasm\cite{Partikian:98,Dayel:99}.

The case of the cell nucleus deserves special consideration.  Here
the primary reactions of importance involve the manipulation of the
genome which is encoded by DNA.  In analogy to our two
reactant model, many of the genome management functions the
cell performs require two specific
portions of the DNA to find each other.  These internal cyclization
reactions may occur between monomers separated by polymer spacers
ranging from less than a persistence length, up to lengths on the
order of the chromosome size\cite{Muller:96,Haber:98}.  If we
identify the polymer segments that flank and bridge the reacting
monomers as crowding particles, then we can immediately generalize
the preceding argument for the rate of crowded reactions to the
internal cyclization of a confined polymer.  Specifically, the
rate of internal cyclization will have a non-monotonic dependence
on the size of the box containing the polymer.  This non-monotonicity
has been previously observed in computer simulations\cite{Abrams:06}.

The onset of crowding limited dynamics at small system sizes leads to
non-monotonic behavior in the relaxation of other structural properties
of the polymer such as the end-to-end vector\cite{Kalb:08}.
This non-monotonicity
is quantitatively different from that of crowded reactions due to the
sub-diffusive behavior of the monomers imposed by the connectivity
constraint.  However, it can be easily explained by
noting that the initial effect
of reducing the volume accessible to the polymer to smaller than
its unconfined size is to reduce the conformational phase space
that the polymer can sample.  This allows the polymer to sample the
phase space faster resulting in shorter relaxation times.  However,
like the free particle case, as the monomer density approaches the
close-packing limit the polymer becomes jammed and the relaxation
times lengthen.

We have shown that a simple, analytically tractable model is able
to quantitatively predict reaction times occurring within crowded
environments.  This model can be generalized to off-lattice systems
and systems with explicit solvent with slight modification.  The
non-monotonicity in the reaction times shown here has broad implications
for reactions within cells as well as the dynamics of confined polymers.

\section*{Acknowledgments}
We would like to thank B. Chakraborty, S. Redner, and A. Grosberg
for useful discussions. This work is supported
by the NSF Grant DMR-0403997.


\begin{thebibliography}{Valens:04}



\bibitem{Fulton:82} A.B. Fulton, Cell {\bf 30}, 345-347 (1982).

\bibitem{Zimmerman:93} S.B. Zimmerman and A.P. Minton, Annu. Rev. Biophys. Biomol. Struct. {\bf 22}, 27 (1993).

\bibitem{Minton:01} A.P. Minton, J. Biol. Chem.{\bf 276}, 10577 (2001).

\bibitem{Zimmerman:91} S.B. Zimmerman and S.O. Trach, J. Mol. Biol. {\bf 222}, 599-620 (1991).

\bibitem{Hall:03} D. Hall and A. P. Minton, Biochem. Biophys Acta {\bf 1649}, 127-139 (2003).

\bibitem{Minton:06} A.P. Minton, J. Cell Sci. {\bf 119}, 2863-2869 (2006).

\bibitem{Berg:83} H.C. Berg {\it Random Walks in Biology}, Princeton University Press  (1983).

\bibitem{Redner:01} S. Redner {\it A Guide to First-Passage Processes}, Cambridge University Press  (2001).

\bibitem{Weeks:00} E. R. Weeks, J. C. Crocker, A. C. Levitt,
Andrew Schofeld, and D. A. Weitz, Science {\bf 287}, 287-631 (2000).

\bibitem{VanBeijeren:85} H. Van Beijeren and R. Kutner, Phys. Rev. Lett. {\bf 55}, 238-241 (1985).

\bibitem{Saxton:87} M.J. Saxton, Biophys. J. {\bf 52}, 989-997 (1987).

\bibitem{Shim:07} J.-u. Shim, G. Cristobal, D.R. Link, T. Thorsen, Y. Jia, K. Piattelli, S. Fraden, J. Am. Chem. Soc. {\bf 129},8825-8835  (2007).

\bibitem{Kao:93} H.P. Kao, J.R. Abney and A.S. Verkman J. Cell Biol. {\bf 120}, 175-184 (1993).

\bibitem{Swaminathan:97} R.Swaminathan, C.P. Hoang, and A.S. Verkman, Biophys. J. {\bf 72}, 1900-1907 (1997).

\bibitem{Elowitz:99} M.B. Elowitz, M.G. Surette, P. Wolf, J.B. Stock and S. Leibler, J. Bacter. {\bf 181}, 197-203 (1999).

\bibitem{Partikian:98} A. Partikian, B. \"{O}lveczky, R. Swaminathan, Y. Li, and A.S. Verkman, J. Cell Biol. {\bf 140}, 821-829 (1998).

\bibitem{Dayel:99} M.J. Dayel, E.F.Y. Hom, and A.S. Verkman, Biophys. J. {\bf 76}, 2843-2851 (1999).

\bibitem{Kasper:98} A. Kasper, E. Bartsch, and H. Sillescu, Langmuir {\bf 14}, 5004-5010 (1998).

\bibitem{Tokuyama:95} M. Tokuyama and I. Oppenheim, Physica A {\bf 216}, 85-119 (1995).

\bibitem{Muller:96}  B. M\"{u}ller-Hill, {\em The Lac Operon: A Short History of a Genetic
Paradigm}, Walter de Gruyter: Berlin, 1996.

\bibitem{Haber:98} J.E. Haber, Annu. Rev. Genet. {\bf 32}, 561-599 (1998).





\bibitem{Abrams:06} C. F. Abrams, N.-K. Lee, and A. Johner,
 Macromolecules, {\bf 39}, 3655 -3663 (2006).  The non-monotonic behaviour
discussed in this reference is not related to that presented in the current
work.  However, Figure 2 shows an increase in the internal cyclization rate
at small system sizes that the authors attribute to ``increased
restriction of monomer mobility''.





\bibitem{Kalb:08} J. Kalb, J.D. Schmit, and B. Chakraborty {\it in preparation }




\end{thebibliography}
\end{document}